\documentstyle[aps,epsfig,prc,12pt]{revtex}
\title{S-wave Pairing of $\Lambda$ Hyperons in Dense Matter}

\author{Shmuel Balberg$^1$ and Nir Barnea$^{1,2}$}
\address{$1$ The Racah Institute of Physics, 
The Hebrew University, Jerusalem 91904, Israel}
\address{$2$ ECT*, European Centre for Theoretical Studies in 
Nuclear Physics and Related Areas,\\ 
Strada delle Tabarelle 286, 1-38050 Villazzano (Trento), Italy}
\date{\today}

\begin{document}
\setcounter{page}{1}

\maketitle
%=================================================================
%    ABSTRACT
%=================================================================
\begin{abstract}
\begin{center}
{\large Abstract}
\end{center}
In this work we calculate the $^1S_0$ gap energies of $\Lambda$ hyperons in 
neutron star matter. The calculation is based on a solution of the BCS 
gap equation for an effective $G$-matrix parameterization of the 
$\Lambda\!-\!\Lambda$ interaction with a nuclear matter background, presented 
recently by Lanskoy and Yamamoto.
We find that a gap energy of a few tenths of MeV is expected 
for $\Lambda$ Fermi momenta up to about $1.3$ fm$^{-1}$. Implications for 
neutron star matter are examined, and suggest the existence of a 
$\Lambda$ $^1S_0$ superfluid between the threshold baryon density for 
$\Lambda$ formation and the baryon density where the $\Lambda$ fraction 
reaches $15\!-\!20\%$.

\end{abstract}

\vskip 2cm


%------------------------------
\setlength{\textheight}{9.0in}
\newpage
\setlength{\topmargin}{-1.3in}
\setlength{\parskip}{0.1in}

%=================================================================
%    INTRODUCTION
%=================================================================

\section{Introduction}

\hspace* {6mm}

The theory of neutron star structure directly relates global properties of 
these stars to various aspects of many-baryon physics. 
One fundamental issue is whether pairing forces among the baryons can 
give rise to baryon superfluids in the inner crust and quantum cores of 
neutron stars. While nucleon pairing in neutron stars has received much 
attention, quantitative estimates of pairing of other baryon species has not 
been performed to date, due to lack of relevant experimental data. 
In this work we use some recent analysis of hypernuclei to make a first 
attempt at determining superfluid gaps for $\Lambda$ hyperons in neutron 
star matter.

Since first suggested by Migdal \cite{Migdal}, nucleon pairing in nuclear 
matter has been the subject of many studies. Both former 
\cite{Taka71,Hoff70} and recent [4-8] works typically find 
$^1S_0$ neutron pairing for neutron matter density, $\rho_n$, in the range 
of $0.1\rho_0\!\leq\rho_n\leq\!0.5\rho_0$, where $\rho_0\!\approx\!0.16$ 
fm$^{-3}$ is the nuclear saturation density. At higher densities, the 
$^1S_0$ interaction turns repulsive, and pairing is possible 
through higher order interactions, mainly $^3P_2$ \cite{Norvg3p2,Taka93}. 
The energy gap found for the $^1S_0$ neutron superfluid is typically 
of the order of a few MeV, although recent works \cite{WAP,Wash} suggest 
that quasi-particle correlations could lower the energy to about 1 MeV. 
Estimates of the $^3P_2$ gap are typically of the order of a few tenths of a 
MeV. It should be noted that published results for the pairing energy gaps 
differ by as much as a factor of three. The difficulty in obtaining 
accurate results is mainly due the the problem of consistently including 
background medium effects. Uncertainties in the two-body 
interactions pose an additional problem. 

As the temperature of a neutron star is expected to drop below 
0.1 MeV ($\sim\!10^9$ K) within about one day from 
its birth, it is widely accepted that nucleon superfluids 
exist in different regions of the star. The qualitative picture of  
a neutron star includes a $^1S_0$ neutron superfluid in its 
inner crust (along with neutron rich nuclei), and a $^3P_2$ neutron 
superfluid in the quantum liquid core. The protons in the core, having a 
partial density of about $10\%$ of the neutrons, are also expected to be in a 
$^1S_0$ superfluid, with an energy gap of about 1 MeV. 
\cite{Wash,Norvg}.  

Baryon superfluids are expected to have a number of important 
consequences on neutron star physics including several observational 
effects, such as pulsar glitch phenomena and cooling rates. 
The crustal neutron superfluid is expected to play a incisive role in 
the driving mechanism of pulsar glitchs, due to pinning of the neutron 
superfluid to the nuclei \cite{Alparal}. Core nucleon superfluids may 
significantly suppress cooling rates that rely on neutrino emission, by 
reducing the available phase space in the final state \cite{SchaabAA,Latal94}.

In this work we focus on the inner core of neutron stars, 
where baryon species other than nucleons are expected to appear. 
It is widely accepted [14-18] that hyperons begin to accumulate at a density 
of about $2\rho_0$, and at a density of $3\rho_0$ the hyperon fraction 
is already about $0.2$. These results are a direct consequence of using 
modern estimates of the interactions of hyperons in nuclear matter, derived 
from hypernuclei experiments. The presence of hyperons has been shown to be of
considerable importance in neutron star cooling rates due to their potential 
to participate in the efficient direct Urca processes. While hyperon direct 
Urca is found to be small compared to the nucleon direct Urca when nucleons 
are non-superfluid, hyperon direct Urca becomes the predominating coolant 
if the nucleons form superfluid pairs 
\cite{Schaabcool}. It is noteworthy that the 
direct Urca mechanism can proceed through hyperon processes for almost 
any hyperon fraction, while the nucleon direct Urca requires a proton 
fraction of at least $0.11\!-\!0.15$ \cite{Lattalcool,Prakcool}. In fact, 
some studies have found that hyperon direct Urca cooling is too rapid to be 
consistent with observed surface temperatures of pulsars 
\cite{Schaabcool,Hanselcool,Nomotocool}. However, if 
hyperons also couple to a superfluid state, as expected for the nucleons,
hyperon direct Urca will also be suppressed, and a large hyperon fraction 
could be easier to coincide with observed cooling rates.  

Hyperon pairing has not been studied previously, as the basic 
obstacles relevant to nucleonic pairing are pronounced for hyperons. 
However, a few measured events in KEK experiments \cite{KEK} attributed to 
doubly-strange $\Lambda\Lambda$ hypernuclei, do offer indication with regard 
to the $\Lambda\!-\!\Lambda$ interaction with a background nuclear matter 
medium. In a recent work Lanskoy and Yamamoto \cite{LanYam} formulated a 
G-matrix parameterization for the $\Lambda\!-\!\Lambda$ interaction, based on 
Nijmegen OBE models. This G-matrix includes a dependence on the density of 
the nuclear matter medium, and reproduces the experimental results of the 
$\Lambda\Lambda$ hypernuclei. 

In this work we aim to employ this formulation to estimate $\Lambda\Lambda$ 
pairing energies in dense matter. We briefly review in Sec. II the formalism 
leading to the gap equation in the $^1S_0$ channel.
The properties of the effective potential used in this work are introduced
in Sec. III. Sec. IV presents our results for the superfluid gap of 
$\Lambda\Lambda$ S-wave pairing in nuclear matter. Implications for 
neutron stars are discussed in Sec. V. Sec. VI contains our conclusions and 
some out-looks regarding hyperon pairing.

%=================================================================
%    SECTION II - Theory (gap equations)
%=================================================================
\section{The Gap Equation}

The BCS theory \cite{BCS} predicts a transition to the superfluid phase
when correlations leading to Cooper  $^1S_0$ pairs give rise to excessive 
binding energy, which overcompensates the increase of energy due to the 
depopulation of the Fermi sea. The appropriate equations have been 
formulated in many works (see, for example, in refs. 
\cite{Norvg85,Wash,Norvg}), and for completeness we review below the main 
results. We note in passing that variation of definitions may lead to 
differences in the numerical coefficients with respect to other works.

The binding energy of a pair with momenta $({\bf k},-{\bf k})$ is found 
through a nonzero solution to the gap equation
\begin{equation} \label{eq:gapeq}
    \Delta_k=-\frac{1}{2} \sum_{k'} {V_{kk'}\frac{\Delta_{k'}}
             {(\xi^2_{k'}+\Delta^2_{k'})^\frac{1}2{}}}  \;\;\; ,
\end{equation}
where $\Delta_k$ is known as the gap function. The potential $V_{kk'}$ is
defined through the matrix element of the $^1S_0$ component of the 
interaction, and $\xi_{k}$ corresponds to the single particle energy, 
$\varepsilon_k$, when measured with respect to the Fermi surface.

Going over to formal integration, the potential term is replaced 
by the potential matrix element: 
$<{\bf k}\!\uparrow-{\bf k}\!\downarrow|
V|{\bf k'}\!\uparrow-{\bf k'}\!\downarrow>$.
In the special case of the $^1S_0$ 
channel the matrix element is independent of the orientation of 
${\bf k}$ and ${\bf k'}$. For a two-particle central potential, $V(r)$, 
the matrix element can be reduced to the form:
\begin{equation} \label{eq:matelm}
    V_{kk'}\equiv<k|V(^1S_0)|k'>={4\pi}
    \int^{\infty}_{0}{r^2\,drj_0(kr)V(r)j_0(k'r)} \;\;\;
\end{equation}
For convenience unit normalization volume is taken for the plane wave 
single particle wave functions; summation over spatial and spin exchange 
terms is implied.

The integral form of the gap equation is thus:
\begin{eqnarray} \label{eq:intgapeq}
    \Delta_k=-\frac{1}{2}\frac{1}{(2\pi)^3}\int{4\pi k'^2\;dk'\;
   V_{kk'}\frac{\Delta_{k'}}{(\xi^2_{k'}+\Delta^2_{k'})^\frac{1}{2}}}= 
   \nonumber  \\
   -\frac{1}{\pi}\int{k'^2\,dk'
   \frac{\Delta_{k'}}{(\xi^2_{k'}+\Delta^2_{k'})^\frac{1}{2}}
   \int r^2\,dr\,j_0(kr)V(r)j_0(k'r)} \;\;\;.
\end{eqnarray}

In this work we use the common ``decoupling approximation'', where the 
Fermi surface is taken to be sharp even in the presence of of the pairing 
correlations. The functions $\xi_k$ are then simply given by 
\begin{equation} \label{eq:defxi_k}
    \xi_{k}=\varepsilon_k-\varepsilon_{k_F}  \;\;\;,
\end{equation}
where we calculated the single particle energies with first order 
Hartree-Fock corrections \cite{deShFesh}. 

The effect of the pairing potential on the single particle energies is 
often characterized by an effective particle mass, $M^*$, which is typically 
lower than the initial (bare) mass by several percent. This mass can be 
estimated through the effective mass approximation:
\begin{equation} \label{eq:defm*}
    M^*=\left(\frac{1}{\hbar^2k_F}
 \frac{d\varepsilon_k}{dk}|_{k=k_F}\right)^{-1}\;\;\;,%\right .|
\end{equation} 
which is usually found to be good up to a few percent  \cite{Wash}.

Note that this effective mass differs from the bulk effective mass, 
found in field theories due to the meson scalar field, also 
typically lower than the bare mass \cite{SMH}. A consistent theory thus 
requires an appropriate ``true'' initial mass, which includes medium effects 
through both the $\Lambda\!-\!\Lambda$ {\it and} 
$\Lambda-$nucleon interactions. However, in the present work we invoke 
a non-relativistic approach, which has no means to consistently combine 
effective bulk masses, and correspondingly 
set the initial mass to be equal to the bare mass, i.e., $M_{\Lambda}=1115.6$ 
MeV (some justification for this may also be found in uncertainties regarding 
values of effective masses at the Fermi surface \cite{MeffFS}). The 
sensitivity of $\Lambda$ pairing to this assumption is examined below.

%=================================================================
%    SECTION III - Yamamoto's potential
%=================================================================
\section{The $\Lambda-\Lambda$ Potential}

In this work we approximate the $^1S_0$ component of the $\Lambda\Lambda$
interaction through a Brueckner $G$-matrix potential. We use the very 
recent $G$-matrix parameterization of Lanskoy and Yamamoto \cite{LanYam} 
derived from the Nijmegen OBE potentials for a $\Lambda\Lambda$ pair in 
nuclear matter. Their evaluation of the $\Lambda\Lambda$ interaction is based 
on measurements of doubly strange hypernuclei observed in experiments 
\cite{KEK}. Analysis of these experiments has suggested both the existence 
of an attractive component in the $\Lambda-\Lambda$ interaction and the 
dependence of this interaction on the properties of the core nucleus
\cite{LLexp}. The strength of the interaction 
is derived from the bond energy of the $\Lambda\Lambda$ pair,
defined as $\Delta B_{\Lambda\Lambda}=B_{\Lambda\Lambda}-2B_{\Lambda}$. 
Here $B_{\Lambda\Lambda}$ is the separation energy of two $\Lambda$'s 
from the nucleus and $B_{\Lambda}$ is the separation energy of a single 
$\Lambda$ from the same nucleus.  

The dependence of the interaction on the nuclear matter density is 
represented in Ref. \cite{LanYam} by a three-range Gaussian form:

\begin{equation} \label{eq:V_LL(r)}
    V_{\Lambda\Lambda}(r)=\sum_{i=1}^{3}
               (a_i+b_ik_{F}(n)+c_ik^2_{F}(n))
               \exp(-r^2/\beta_{i}^{2}) \;\;\; ,
\end{equation}
where $k_F(n)$ is the nucleon Fermi momentum. Assuming symmetric nuclear 
matter (as is the case for light hypernuclei), $k_F$ is related to the 
nuclear density $\rho_N$ by $k_F\!=\!(3\pi^2\frac{1}{2}\rho_N)^{\frac{1}{3}}$. 
Since the $\Lambda$ is an isospin singlet, it also seems safe to
apply equation (\ref{eq:V_LL(r)}) to non-symmetric nuclear matter
of density $\rho_N$.

The ranges $\beta_i$ and the strength parameters $a_i,b_i,c_i$ are taken 
from model ND of \cite{LanYam} and are listed in table \ref{V_LL NS}. This 
model successfully reproduces the experimental result of 
$\Delta B_{\Lambda\Lambda}=4.9\pm0.7$ MeV of $^{13}_{\Lambda\Lambda}B$ 
\cite{KEK}.

The radial dependence of the $\Lambda\!-\!\Lambda$ interaction is demonstrated 
in Fig. 1 which shows $V_{\Lambda\Lambda}(r)$ for nuclear matter densities
of $\rho_N/\rho_0\!=\!1$, $\rho_N/\rho_0\!=\!2.5$ and $\rho_N/\rho_0\!=\!5$. 
At short distances the interaction is always repulsive, reflecting the core 
repulsion of the bare interaction (we note that $G$-matrix approximations 
typically yield soft cores \cite{WAP} which substitute the need for short 
range cut-off necessary in other interaction models). At intermediate 
distances the $^1S_0$ yields an attractive force of several tens of MeV's, 
which is strong enough to yield the pairing of the superfluid state.

As can be seen in Fig. 1, the dependence of the interaction on the nuclear 
matter density is rather weak. This implies that the existence of $^1S_0$ 
pairing should have only a moderate dependence on the density of the nuclear 
matter medium. We note, however, that the magnitude of the interaction tends 
to grow larger for a larger background density.

It must be noted that the $G$-matrix parameterization is fitted to 
match experimental results for
different nuclei, and is thus likely to be valid for a nuclear matter 
background with a density of $\rho_N\!\approx\!\rho_0$. In the following 
analysis we assume that the $G$-matrix is valid for higher densities as 
well. Clearly this is a somewhat crude assumption, especially since 
the $G$-matrix does not incorporate any relativistic effects 
which could be significant at densities relevant to neutron star 
cores ($\rho\!\geq\!2\rho_0$). Hence, the results derived below must be 
viewed as preliminary estimates. More founded results must await better 
established $\Lambda\Lambda$ potentials in high density nuclear matter.

%=================================================================
%    SECTION IV - Results for $\Lambda$ pairing in dense matter
%=================================================================
\section{Results for $\Lambda$ pairing in dense matter}

Using the $\Lambda\Lambda$ potential described in the previous Section, 
we have solved the gap equation, Eq. (\ref{eq:intgapeq}), for $\Lambda$ 
hyperons in a nuclear matter background. The solution is found by iterations, 
when the integration is performed with a few hundred integration points, 
exponentially spaced around $k_F(\Lambda)$. The exponential spacing is 
required since the integrand in Eq. (\ref{eq:intgapeq}) is sharply peaked at 
the Fermi momentum. This behavior is demonstrated in Fig. 2 which shows the
integrand for a $\Lambda$ Fermi momenta of $k_F(\Lambda)\!=\!1.0$ 
fm$^{-1}$ and nuclear matter densities of $\rho_N\!=\!2.5\rho_0$ and 
$\rho_N\!=\!5.0\rho_0$. Fig. 2 also indicates the need for a large cutoff 
momenta in the calculation of the gap function in Eq. (\ref{eq:intgapeq}).

Solution of the gap equations gives the gap function $\Delta_k$ for any 
combination of values for the nuclear matter background density
and the $\Lambda$ Fermi momenta. Fig. 3 shows the gap function
for the same values of $\rho_N$ and $k_F(\Lambda)$ as in Fig. 2. The gap 
function falls off from its maximum at $k(\Lambda)\!=\!0$, and 
varies very rapidly around $k_F(\Lambda)$. The gap function is also found 
to be negative over a wide range of higher momenta. As can be seen in Fig. 3, 
the gap energy $\Delta_k$ is always larger in absolute magnitude for 
larger nuclear matter density. 
This results from the enhancement of the two-particle interaction at higher 
background densities, as seen in Fig. 1. Since the size of the superfluid gap 
for a given $k_F(\Lambda)$ is mostly dependent on the two-particle 
interaction at distances of about $1/k_F(\Lambda)$, the gap energy grows 
larger for a larger density of the background nuclear medium.
Correspondingly, The integrand of Eq. (\ref{eq:intgapeq}) (Fig. 2) also 
increases for a larger $\rho_N$. 

We note that qualitative results such as those shown in Figures 2-3 are common 
also in solutions of the gap equations for nucleons \cite{Wash,Norvg}.
The need for a large cutoff momenta is of particular 
importance, since it clarifies why the gap energy estimated through 
the weak-coupling-approximation (WCA) \cite{LifPit} systematically 
underestimates the gap energy. In this approximation one essentially assumes 
that it is sufficient to integrate Eq. (\ref{eq:intgapeq}) over a narrow range 
near $k_F$. Indeed, gap energies found for nucleons through the WCA are 
usually lower by a factor of two and more than those derived by a self 
consistent solution of the gap equation 
(see, for example, in Ref. \cite{Wash}).

The prevalent result of the solution of the gap equation is 
the value for the gap energy at the Fermi surface, 
$\Delta_F\!\equiv\!\Delta_{k_F}$. The resulting function $\Delta_F(k_F)$
has a typical bell shape, ranging from $\Delta_0\!=\!0$ to some maximum 
value and then falling off again to zero. This later part of the 
$\Delta_F(k_F)$ arises from the decrease of mean inter-particle distance at
higher $k_F$, as the $\Lambda$'s sample more of the repulsive core. This 
physical mechanism causes the S-wave superfluidity to vanish at large 
$\Lambda$ partial densities. 
The $\Delta_F(k_F(\Lambda))$ dependence for nuclear matter background 
densities equal to $2\rho_0,2.5\rho_0,3\rho_0$ and $5\rho_0$ are shown in 
Fig. 4. The corresponding values of the gap energies and the effective 
$\Lambda$ masses for $\rho_N=2.5\rho_0$ and $\rho_N=5.0\rho_0$ are given in 
Table 2.

As is expected from Figs. 2-3, the gap energy for a given $k_F(\Lambda)$ 
increases along with the density of the nuclear matter background, $\rho_N$. 
However, for matter composed of nucleons and $\Lambda$'s, increasing 
$\rho_N$ alone corresponds to lowering the fraction of the $\Lambda$'s. 
On the other hand keeping the $\Lambda$ fraction constant while increasing the 
total density amounts to an increase of $\Lambda$ Fermi momenta, and it is 
clear from Fig. 4 that increasing $k_F(\Lambda)$ beyond 
$0.8$ fm$^{-1}$ should lead to a decline in the gap energy. Thus, 
increasing the total baryon density with a given $\Lambda$ fraction tends to 
reduce the gap energy, while a larger total baryon density 
also means a larger nuclear matter density, which should increase the gap 
energy. Hence, these two trends compete when the total 
baryon fraction is increased and the $\Lambda$ fraction is kept constant.

In Fig. 5 we compare these two trends by presenting gap energies 
at the Fermi surface as a function of 
the total baryon density, $\rho_B$, of matter composed of nucleons 
and $\Lambda$'s. The curves represent constant $\Lambda$ fractions of 
$5\%$, $10\%$, $15\%$ and $20\%$ of the total baryon population. 
As it happens for the particular pairing interaction used in this work, 
the two trends balance for a $\Lambda$ fraction of about $5\%$, and for 
a larger $\Lambda$ fraction the gap energy decreases to 
zero as the total baryon density is increased. These results have direct 
implications on the gap energies in neutron star matter, as is discussed in 
the next Section.

We now return to the problem of the ``true'' effective masses of the baryons 
in dense matter. So far we have assumed that the initial mass of the 
$\Lambda$ hyperons on the Fermi surface is equal to the bare mass, 
$M_{\Lambda}\!=\!1115.6$ MeV. Since we do not combine a self consistent 
treatment of bulk effects and the relativistic properties of the 
interactions, we must resort to arbitrary parameterization to examine the 
dependence of the pairing energies on the initial mass. A more accurate 
derivation of consistent interactions and masses is deferred to future work.

Fig. 6 demonstrates the dependence of the gap energies on the ``true'' 
effective mass of the $\Lambda$ hyperons in the matter. The results shown are 
for a nuclear matter density of $\rho_N\!=\!2.5\rho_0$ and $\Lambda$ 
initial masses taken as 0.7, 0.85 and 1.0 times the bare mass (as mentioned 
in Sec. II, the effective mass derived by the solution to the superfluid 
equations is always lower than the initial one by several percent). 
As is expected, a lower mass leads to higher single-particle energies 
for any given momenta, and this yields lower gap energies. However, the 
basic existence of a superfluid gap of $\Delta_F\!\geq\!0.1$ MeV for 
$k_F(\Lambda)\!\leq\!1.3$ fm$^{-1}$ is found also for effective masses 
lower than the bare mass.

A final point of interest is the the dependence of the gap energies on the 
matter temperature. The importance attributed to this dependence is of obvious 
in view of the implications of baryon superfluidity on neutron star cooling 
rates. We hereby follow the approach of Elgar$\o$y et al. \cite{Norvg} in 
estimating this dependence for $\Lambda\Lambda$.

The gap equation at a finite temperature $T$ is given by revising Eq. 
(\ref{eq:intgapeq}) to the form:
\begin{equation} \label{eq:gapT}
 \Delta_k(T)=-\frac{1}{2}\frac{1}{(2\pi)^3}\int_{0}^{\infty}{4\pi k'^2\,dk'\,
    V_{kk'}\frac{\Delta_{k'}(T)}
    {(\xi^2_{k'}(T)+\Delta^2_{k'}(T))^\frac{1}{2}}}
    \tanh\left(\frac{(\xi^2_{k'}(T)+\Delta^2_{k'}(T))^\frac{1}{2}}
    {2k_BT}\right),
\end{equation}  
where $k_B$ is the Boltzmann constant. We solve Eq. (\ref{eq:gapT}), 
while approximating the single-particle energies to be ``frozen'', i.e.,
assuming that $\xi_{k'}(T)\!=\!\xi_{k'}(0)$. This should be a reasonable 
approximation for neutron stars, since the temperature range of interest 
is much lower than the Fermi energy (see \cite{Norvg} and references therein).
We also assume that the two-particle interaction is not sensitive to 
the temperature in the range of interest. The gap equation is then solved 
in similar fashion to the zero-temperature case.

The temperature dependence of the gap energy at the 
Fermi surface, $\Delta_F(T)$ for background nuclear matter density of 
$\rho_N\!=\!2.5\rho_0$ and $\rho_N\!=\!5\rho_0$ is shown in Fig. 7. 
Also shown are the critical temperatures, $T_c$, estimated from the 
WCA, given by 
\cite{LifPit}
\begin{equation} \label{eq:WCAgapT}
    k_BT_c\approx0.57\Delta_F(T=0) \;\;\; .
\end{equation}

As in the case of nuclear matter \cite{Norvg}, we see  that 
the WCA does yield good agreement with the results of the full solution, 
provided that the value of $\Delta_F(T=0)$ is taken from the gap equation 
solution rather than the WCA for the gap, as explained above. 

%=================================================================
%    SECTION V - Implications for neutron star matter
%=================================================================
\section{Implications for neutron star matter}

Modern estimates [14-18] of hyperon formation in neutron stars agree 
that hyperons begin to accumulate in neutron star matter at baryon densities 
of about $2\rho_0$. In particular, the threshold baryon density for 
$\Lambda$ formation is found to be about $2.5\rho_0$, when the chemical 
potential of the neutrons grows large enough to compensate for the mass 
difference $M_{\Lambda}-M_n$. While the fine details of the $\Lambda$ 
fraction in the matter are model dependent, these basic features are 
widely accepted. We stress that this consensus is an immediate result of 
employing realistic values for the interaction of $\Lambda$ hyperons in 
nuclear matter, based on experimental data of $\Lambda$-hypernuclei 
\cite{hypnucrev}.

An example of the equilibrium composition of neutron star matter 
(assuming $T=0$) is given in Fig. 8a, based on an equation of state similar 
to the $\delta\!=\!\gamma\!=\!\frac{5}{3}$ model of Ref. \cite{BGH}.
The steep rise in the $\Lambda$ fraction when they first appear in the matter 
is common to all works that examined hyperon formation in neutron stars. 
This behavior is caused by the fact that lowering the nucleon fraction lowers
the nucleon-nucleon repulsion and the nucleon Fermi energies, while the net 
interaction among the $\Lambda$'s is still attractive. Eventually the 
$\Lambda$ fraction saturates, typically at $0.1-0.2$, and 
continues to grow slowly up to as much as 0.3 at higher densities.

Recent theoretical and experimental results of $\Sigma^-$-atoms suggest that 
the interaction of $\Sigma$ hyperons in nuclear matter includes a strong 
isoscalar repulsive component \cite{SigAtom}. If such repulsion exists, 
formation of $\Sigma$ hyperons in neutron star matter is suppressed 
\cite{SBH,BGH}, and $\Lambda$ production in the matter is somewhat enhanced, 
both by a lower threshold density and by a sharper rise of the 
$\Lambda$ fraction. The main effect, though, is the formation of $\Xi^-$ 
hyperons which begins at significantly lower densities (about $3\rho_0$), 
providing the favorable negatively charged baryon fraction. The equilibrium 
compositions of matter without $\Sigma$'s is shown in Fig. 8b, 
using an equation of state otherwise identical to that of Fig. 8a. 

In view of the absence of any experimental data on medium effects regarding 
different hyperon species, we assume in the following analysis that the 
pairing interaction discussed in Sec. 3 is valid also for a background 
matter which includes other species besides nucleons (i.e. $\Sigma$ and 
$\Xi$ hyperons). For densities up to $\sim\!5\rho_0$ this is a reasonable 
assumption, since the non-$\Lambda$ matter is highly dominated by the 
nucleons. Thus, for every combination of the total baryon density and particle 
fractions we take $\rho_{bg}\!\equiv\!\rho_B\!-\!\rho_{\Lambda}$ as the 
background density $\rho_N$ for the calculation of the $^1S_0$ gap energy.

The $\Lambda\Lambda$ gap energies found for the baryon compositions 
of Figs. 8a-8b are shown in Fig. 9, as a function of the total baryon 
density. Also shown are the gap energies for the equilibrium composition for 
model PLZ of Schaffner and Mishustin \cite{SMH}, which predict $\Lambda$ 
accumulation at slightly lower densities than the equations of \cite{BGH}.

As seen in Fig. 9, the qualitative behavior of the gap energies is common to 
all three equations of state. Pairing to a superfluid state essentially 
takes place once the $\Lambda$'s appear in the matter, and rises sharply to 
a maximum value following the sharp rise of the $\Lambda$ fraction in the 
matter. The partial density of the non-$\Lambda$ baryons is almost constant 
in this range of total baryon densities. Hence, the curves approximately 
follows the gap energy dependence on $k_F(\Lambda)$ for a given background 
density, as shown in Fig. 4. 
The pairing energy rises sharply as the total baryon density is increased, 
and then, as is expected from Fig. 5, begins to decline once the $\Lambda$ 
fraction exceeds about $0.05$, 
(note that for $\rho_B\!=\!2\rho_0$, $k_F(\Lambda)\!\approx\!0.8$ fm$^{-1}$ is 
reached when the $\Lambda$ fraction is about $5\%$). 
Since the $\Lambda$ fraction begins to saturate at a value of 
$0.1\!-\!0.2$, the decline of the gap energy is not as steep 
as in the rising part. The rate of this decline is thus 
somewhat model dependent, particularly whether other hyperon species 
(i.e. the $\Sigma^-$) compete with $\Lambda$ formation.

It should be noted that these results are qualitatively similar to those found 
for proton $^1S_0$ pairing in neutron star matter, where protons are a minority 
among the nucleons. The density range found for a superconducting proton state 
lies between the threshold for free proton appearance up to 
densities where the proton fraction reaches about $0.1\!-\!0.2$ 
\cite{Wash,Norvg}. However, since the proton fraction in neutron star matter 
is expected to rise much more moderately as a function of the total baryon 
density than the $\Lambda$ fraction (see Figs. 8a-8b), the density range 
where a proton superconductor exists is typically larger than that 
found here for the $\Lambda$ superfluid.  

%=================================================================
%    SECTION VI - Conclusions and Outlook
%=================================================================
\section{Conclusions and Outlook}

In this work the $^1S_0$ pairing energy of $\Lambda$ hyperons in a nuclear 
matter background was evaluated using the $G$-Matrix effective 
interaction presented by Lanskoy and Yamamoto \cite{LanYam}. 
We find that a gap energy of a few tenths of MeV is expected 
for a $\Lambda$ Fermi momenta, $k_F(\Lambda)$, below 1.3 fm$^{-1}$. The gap 
energy is dependent both on the $\Lambda$ Fermi momenta and on the 
density of the background nuclear matter, $\rho_N$. For 
$\rho_N\!\geq\!2\rho_0$ the gap energy 
for a given $k_F(\Lambda)$ increases with increasing $\rho_N$. 

Employing these results to neutron star matter with hyperons yields 
$\Lambda\Lambda$ $^1S_0$ pairing for a baryon density range between the 
threshold density for $\Lambda$ appearance to about the baryon density where
the $\Lambda$ fraction reaches $\sim0.2$. A maximum gap energy of 
$0.8\!-\!0.9$ MeV is achieved for a $\Lambda$ fraction of about $0.05$.  
While the exact range of densities where such pairing exists is 
model-dependent, the qualitative picture seems to be common to all equations 
of state which are based on modern evaluations of the $\Lambda\!-\!$nucleon 
interaction in nuclear matter. Gap energies in this range are larger than 
the temperature predicted in neutron star cores, and thus imply that a 
$\Lambda$ $^1S_0$ superfluid will exist in the core, typically within a 
baryon density range of $\rho_B\!\approx\!2\!-\!3\rho_0$. 

We comment that the present results must be treated as a 
preliminary evaluation of $\Lambda$ 
pairing in dense matter. The evaluation of the two-particle interaction is 
based on hypernuclei experiments, where the nuclear matter density is 
limited to $\rho_N\!\approx\!\rho_0$, so that the effective interaction might 
not be as good an approximation as in the case of neutron pairing. 
In particular, the present work does not include 
relativistic corrections which might be significant at the baryon densities 
where hyperons form in neutron star matter (note however, that relativistic 
corrections for proton $^1S_0$ pairing at about the same densities have been 
found to introduce only small corrections to the nonrelativistic results 
\cite{NorvgPRL}). It is also noteworthy that we have not included 
particle-hole correlations which have been shown to be important in the 
evaluation of gap energies \cite{WAP}. In short, further work is necessary 
to produce more realistic results, preferably with a better founded 
$\Lambda\!-\!\Lambda$ interaction in a high density nuclear matter background.

We believe that formal treatment of the non-nucleon component in the 
background neutron star matter, will not significantly effect 
the results found here. This is especially true if $\Sigma$ hyperon 
formation is suppressed, so that the baryon equilibrium compositions include 
only nucleons and $\Lambda$'s throughout the entire range where 
pairing is expected. Nonetheless, taking other hyperon species into account 
is clearly desirable in a more rigorous model. Obviously, hyperon-hyperon 
interactions in a dense matter background will also provide a basis for 
estimation of possible pairing of other hyperon species in neutron stars. 
For example, $\Sigma^-$ pairing is of special interest, since the $\Sigma^-$ 
is also expected to appear at relatively low baryon densities in neutron stars 
(if $\Sigma$ formation is not suppressed). However, no relevant experimental 
data is currently available. The commonly assumed universal hyperon-hyperon 
interaction implies that the $\Lambda\Lambda$ gaps may serve as indication 
for $\Sigma$ and $\Xi$ pairing in dense matter. More accurate results 
require, however, the inclusion of isospin-dependent forces, which are absent 
in the $\Lambda$ case.

The large majority of dense matter equations of state require 
neutron star central densities larger than the threshold density for $\Lambda$ 
formation, i.e. baryon densities larger than $\sim\!2.5\rho_0$. Hence, it is 
likely that neutron stars do include a region where the $\Lambda$'s 
pair to a $^1S_0$ superfluid. Whether or not the central density of a 
neutron star exceeds the density range for $^1S_0$ $\Lambda\Lambda$ pairing 
depends on its mass and on the actual equation of state. Note, however, 
that at larger densities higher order pairing may also be available, 
including inter-species pairing \cite{Taka93}. In fact, 
$\Lambda n$ pairing may be more likely than $pn$ pairing, since at baryon 
densities of $\rho_B\!\geq\!\sim\!4\rho_0$ 
the $\Lambda$ and neutron fractions are expected to be comparable.

Finally, we recall that the existence of a $^1S_0$ $\Lambda$ superfluid for 
baryon densities relevant to neutron stars implies significant suppression of 
$\Lambda$-direct Urca cooling. The onset of superfluidity reduces the 
neutrino emissivity, along with the heat capacity and thermal conductivity, 
by a factor of $\exp(-\Delta_F/k_B T)$. In view of our results here, 
we suggest that implications of hyperon superfluidity on neutron star cooling 
rates are well worth examination. 

%{\large Acknowledgments} 
\acknowledgements
 
We are grateful to Avraham Gal for valuable discussions and comments. 
We also thank J\"{u}rgen Schaffner for providing us with equilibrium 
compositions of the relativistic mean field calculations. 
One of us (S.B.) also thanks Meir Weger for helpful 
guidence regarding the BCS theory. This research was 
partially supported by the U.S.-Israel Binational Science Foundation 
grant 94-68. 

\newpage

%\begin{thebibliography}{99}

\pagebreak

\begin{center}
{\large Figure Captions}
\end{center}
\vskip -0.5cm
\setlength{\parskip}{0.1in}

Figure 1: The radial dependence derived for the $\Lambda\Lambda$ $G$-matrix 
interaction presented in \cite{LanYam}. The curves correspond to nuclear 
matter background densities of $\rho_N\!=\!\rho_0$, $2.5\rho_0$ and 
$5\rho_0$, where $\rho_0$ is the nuclear saturation density.

Figure 2: The integrand of the gap equation, Eq. (\ref{eq:intgapeq}), for 
$k_F(\Lambda)\!=\!1$ fm$^{-1}$, as a function of the secondary momenta $k'$. 
The curves correspond to nuclear matter background densities of 
$\rho_N\!=\!2.5\rho_0$ and $5\rho_0$.

Figure 3: The gap function, $\Delta_k$, when 
$k_F(\Lambda)\!=\!1$ fm$^{-1}$ and nuclear 
matter background densities of $\rho_N\!=\!2.5\rho_0$ and $5\rho_0$.

Figure 4: The gap energy $\Delta_F$ for $\Lambda\Lambda$ pairing as a function
of the Fermi momenta, for nuclear matter background densities of 
$\rho_N\!=\!2\rho_0$, $2.5\rho_0$, $3\rho_0$ and $5\rho_0$.

Figure 5: The gap energy for $\Lambda\Lambda$ pairing as a 
function of the total baryon density, $\rho_B$, for different fixed 
$\Lambda$ fractions.

Figure 6: The gap energy $\Delta_F$ for $\Lambda\Lambda$ pairing as a function
of the Fermi momenta, for different values of the initial 
mass of $\Lambda$ hyperons in the matter. The nuclear matter background 
density is taken as $\rho_N\!=\!2.5\rho_0$.

Figure 7: Temperature dependence of the gap energy for 
$k_F(\Lambda)\!=\!1$ fm$^{-1}$ for nuclear matter background densities of 
$\rho_N\!=\!2.5\rho_0$ and $5\rho_0$. Also indicated are the corresponding 
weak-coupling estimates for the critical temperatures.

Figure 8a: The equilibrium compositions of neutron star matter with hyperons, 
as a function of the total baryon density, $\rho_B$. The compositions were 
calculated with an equation of state similar to the 
$\delta\!=\!\gamma\!=\!\frac{5}{3}$ model from \cite{BGH}.

Figure 8b: Same as Fig. 8a, but when $\Sigma$ 
hyperons are repelled by the nucleons and their formation is thus suppressed.

Figure 9: The gap energy of $\Lambda\Lambda$ $^1S_0$ pairing 
in neutron star matter as a function of the total baryon density. The 
equilibrium compositions of the matter are those of Figs. 8a (BG+$\Sigma$) 
and 8b (BG-$\Sigma$), and for model PLZ of Schaffner and Mishustin 
\cite{SMH} (SM$_{PLZ}$). 

\pagebreak
% TABLES

%==============================================================================
\begin{table} 
\caption{\label{V_LL NS}} Parameters of the $^1S_0$ state of the 
$\Lambda\Lambda$ $G$-matrix potential (model ND of \cite{LanYam}) 
\begin{center}
\begin{tabular}{c c c c}
$\beta_i$ & $a_i$     & $b_i$     & $c_i$         \\
 (fm)     & (MeV)     & (MeV fm)  & (MeV fm$^2$)  \\ \hline
$0.5$     & $ 835.5$  & $-252.7$  & $122.7 $      \\
$0.9$     & $-298.5$  & $ 156.6$  & $-55.07$      \\
$1.5$     & $-10.80$  & $3.0398$  & $-1.126$      \\
\end{tabular}
\end{center}
\end{table}
%========================================================================

%==============================================================================
\begin{table} 
\caption{\label{D_FL}} $\Lambda\Lambda$ $^1S_0$ pairing energy gaps
and effective masses 
\begin{center}
\begin{tabular}{c c c c c}
\multicolumn{1}{c}{ } &
\multicolumn{2}{c}{$\rho_N=2.5\rho_0$} &
\multicolumn{2}{c}{$\rho_N=5\rho_0$}\\
$k_F(\Lambda)$ & $M^*/M$   & $\Delta_F$  & $M^*/M$    & $\Delta_F$ \\
(fm$^{-1}$)    &           &   (MeV)     &            &   (MeV)    \\ \hline
$0.2$          & $ 0.9967$ & $ 0.0432 $  & $ 0.9963 $ & $ 0.1321 $   \\
$0.3$          & $ 0.9895$ & $ 0.1767 $  & $ 0.9881 $ & $ 0.3897 $     \\
$0.4$          & $ 0.9771$ & $ 0.3749 $  & $ 0.9740 $ & $ 0.7143 $     \\
$0.5$          & $ 0.9596$ & $ 0.5868 $  & $ 0.9543 $ & $ 1.0371 $     \\
$0.6$          & $ 0.9383$ & $ 0.7628 $  & $ 0.9304 $ & $ 1.2998 $     \\
$0.7$          & $ 0.9150$ & $ 0.8677 $  & $ 0.9042 $ & $ 1.4611 $     \\
$0.8$          & $ 0.8915$ & $ 0.8735 $  & $ 0.8779 $ & $ 1.4937 $     \\
$0.9$          & $ 0.8693$ & $ 0.7826 $  & $ 0.8532 $ & $ 1.3876 $     \\
$1.0$          & $ 0.8497$ & $ 0.6130 $  & $ 0.8313 $ & $ 1.1574 $     \\
$1.1$          & $ 0.8335$ & $ 0.4027 $  & $ 0.8132 $ & $ 0.8409 $     \\
$1.2$          & $ 0.8211$ & $ 0.2053 $  & $ 0.7993 $ & $ 0.5262 $     \\
$1.3$          & $ 0.8128$ & $ 0.0495 $  & $ 0.7900 $ & $ 0.1810 $ \\ 
\end{tabular}
\end{center}
\end{table}
%========================================================================
%\end{document}

\newpage
\epsfig{file=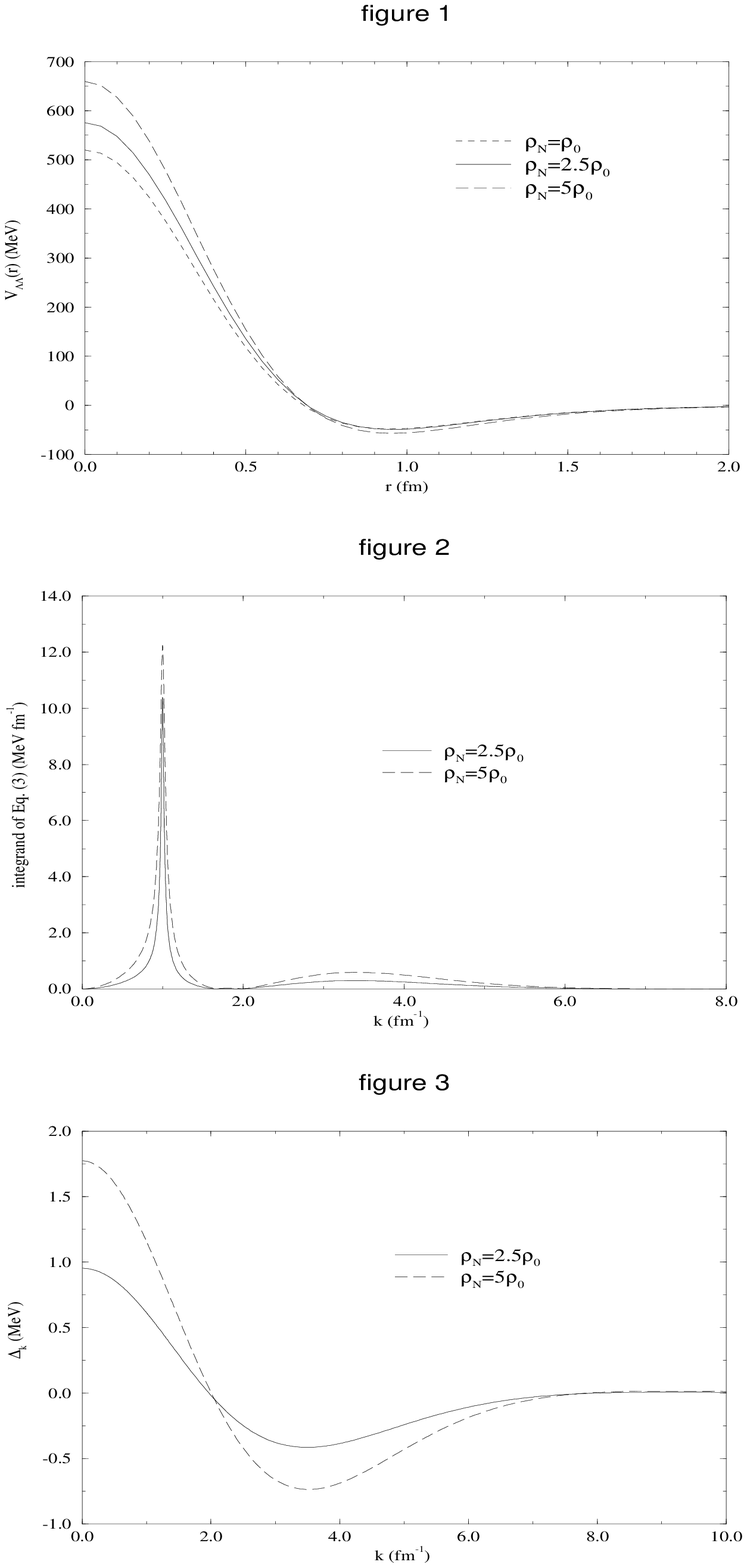,height=23.0cm,width=15.5cm}
\newpage
\epsfig{file=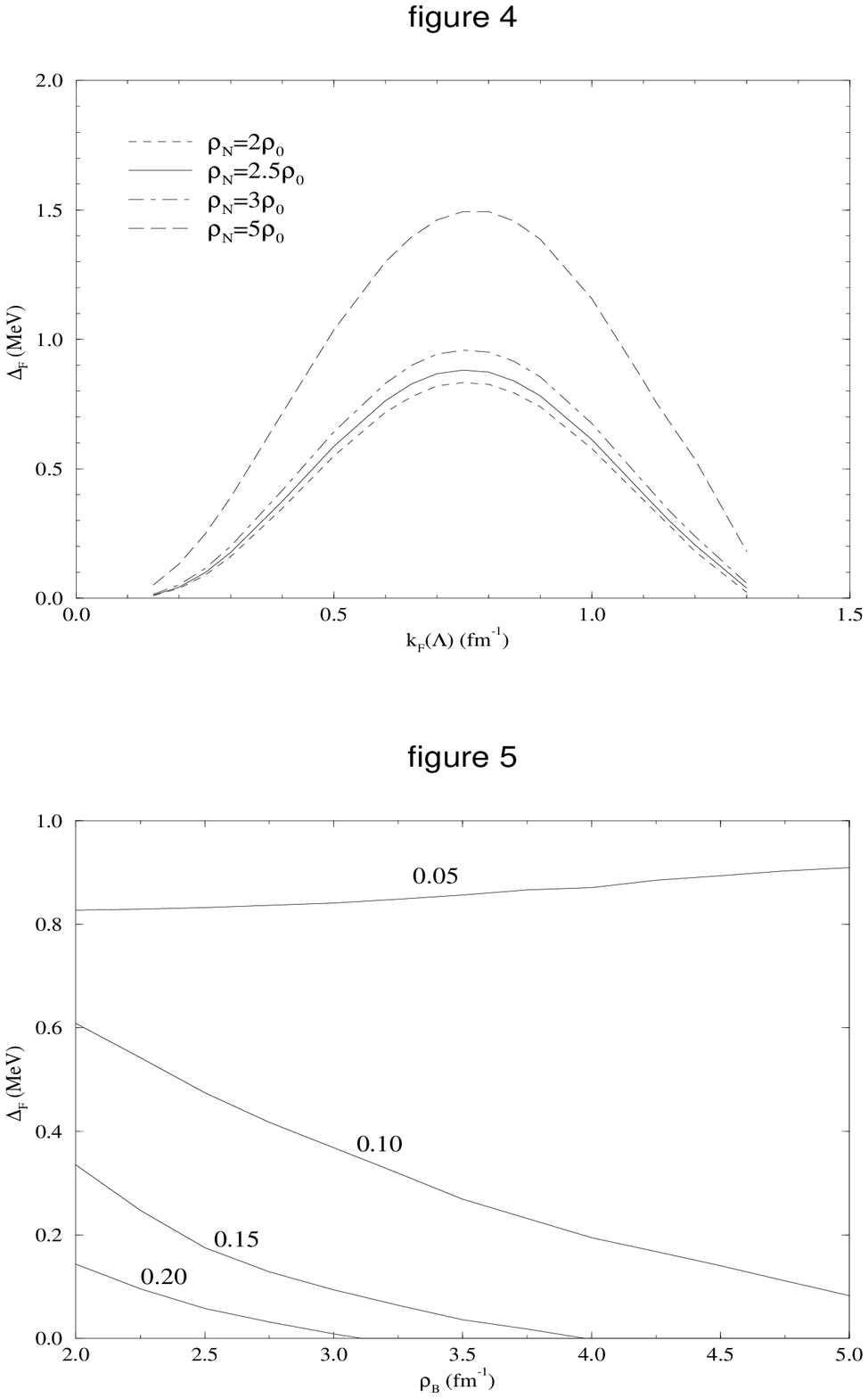,height=23.0cm,width=15.5cm}
\newpage
\epsfig{file=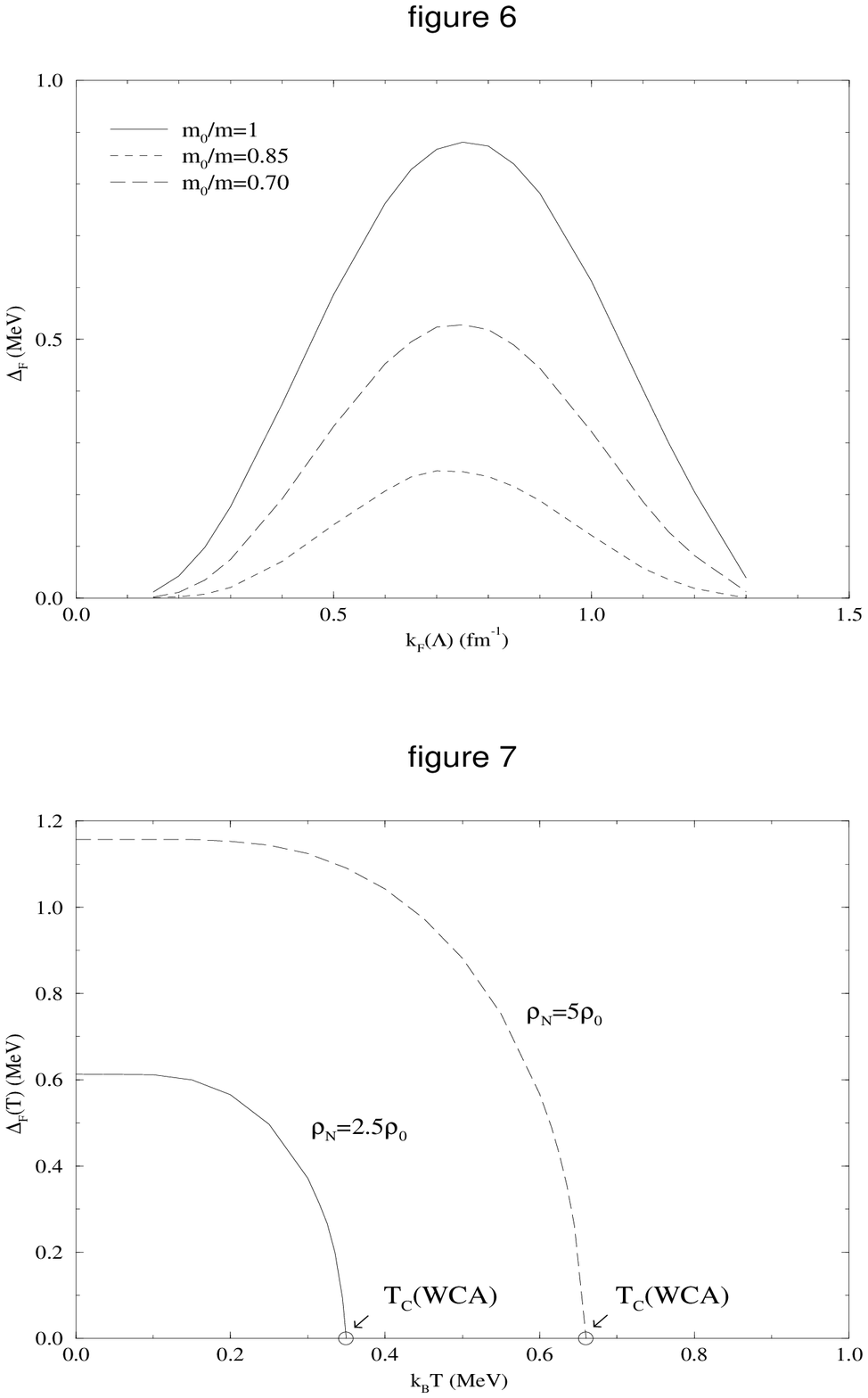,height=23.0cm,width=15.5cm}
\newpage
\epsfig{file=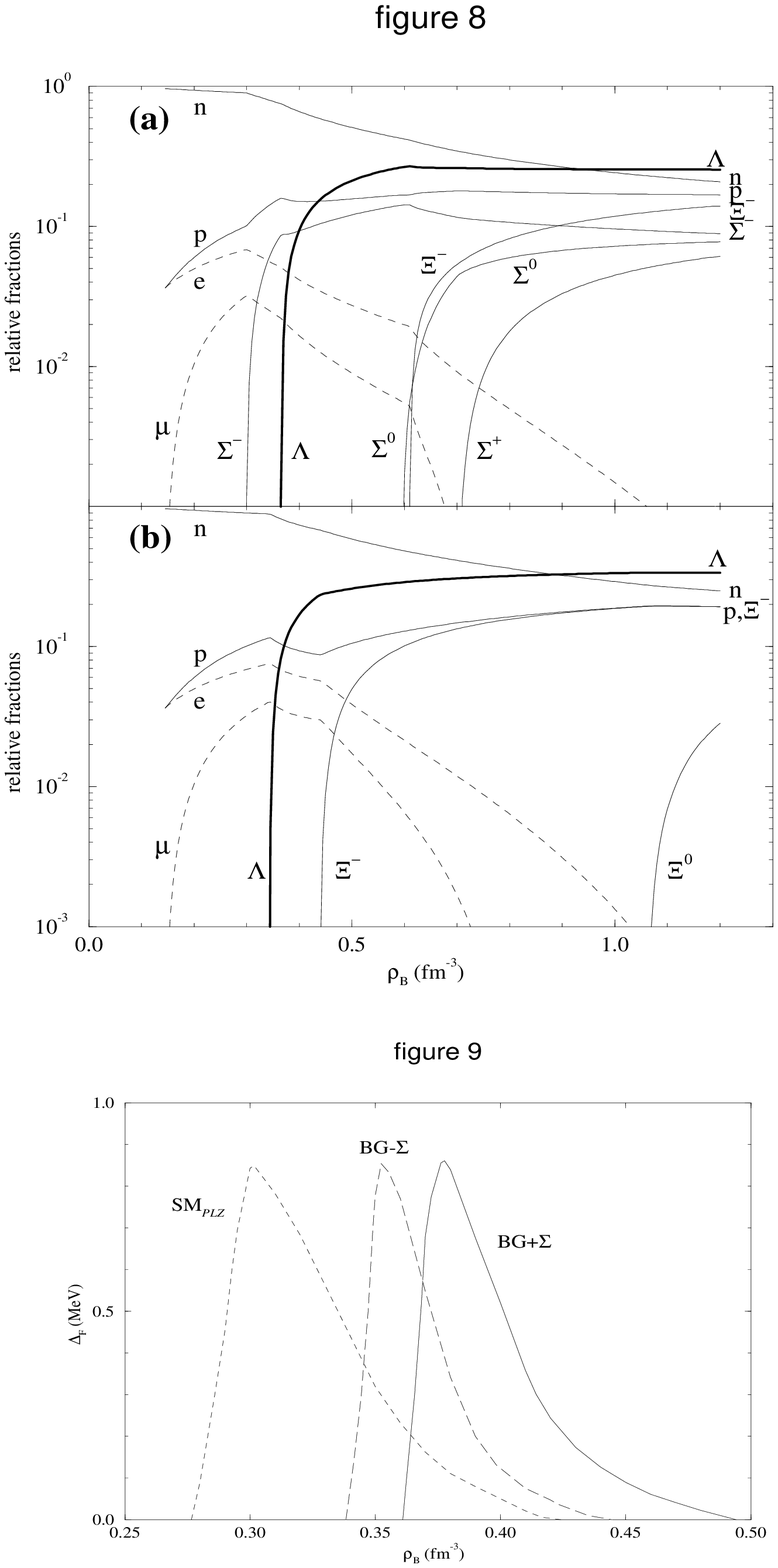,height=23.0cm,width=15.5cm}
\end{document}